# Flat Bands in Buckled Graphene Superlattices

Yuhang Jiang[1], Miša Anđelković[2], Slaviša P. Milovanović[2], Lucian Covaci[2], Xinyuan Lai[1], Yang Cao[3,4], Kenji Watanabe[5], Takashi Taniguchi[5], Francois M. Peeters[2], Andre K. Geim[3,4] and Eva Y. Andrei[1]

[1]Department of Physics and Astronomy, Rutgers University, 136 Frelinghuysen Road, Piscataway, NJ 08855 USA

[2]Departement Fysica, Universiteit Antwerpen, Groenenborgerlaan 171, B-2020 Antwerpen, Belgium

[3]School of Physics and Astronomy, University of Manchester, Oxford Road, Manchester M13 9PL, United Kingdom

[4]Manchester Centre for Mesoscience and Nanotechnology, University of Manchester, Oxford Road, Manchester M13 9PL, United Kingdom

[5]Advanced Materials Laboratory, National Institute for Materials Science, 1-1 Namiki, Tsukuba 305-0044, Japan

**Interactions between stacked two-dimensional (2D) atomic crystals can radically change their properties, leading to essentially new materials in terms of the electronic structure. Here we show that monolayers placed on an atomically flat substrate can be forced to undergo a buckling transition, which results in periodically strained superlattices. By using scanning tunneling microscopy and spectroscopy and support from numerical simulations, we show that such lateral superlattices in graphene lead to a periodically modulated pseudo-magnetic field, which in turn creates a post-graphene material with flat electronic bands. The described approach of controllable buckling of 2D crystals offers a venue for creating other superlattice systems and, in particular, for exploring interaction phenomena characteristic of flat bands.**

Buckling transitions in thin stiff membranes including graphene are ubiquitous. They generally occur when in-plane compressive strain causes a periodic out-of-plane deformation of the membrane[1-3]. The compression can be generated by thermal cycling, processing in the presence of solvents or substrate-induced stress. Upon exceeding a critical strain, the membrane can reduce its elastic energy through out of plane distortions resulting in intriguing periodic 1D or 2D patterns (Fig. 1) whose specific structure is dictated by boundary conditions, geometry and strain distribution[3]. In graphene membranes, the buckling can produce a periodic pseudo-magnetic field (PMF)[4-7] which reconstructs the low energy band structure into a series of flat bands[8-10]. The quenched kinetic energy in flat bands promotes electron-electron interactions and facilitates the emergence of strongly correlated electronic phases that are expected to survive up to high temperatures when the flat bands take up a large fraction of the Brillouin zone[11-13]. A celebrated example is the sequence of the magnetically generated flat energy bands, Landau levels. Bringing the Fermi energy within a Landau level leads to correlated phases including fractional quantum-Hall states[14-17] and the Wigner crystal[18]. Magnetically-induced flat bands have however limited applicability because the broken time reversal symmetry precludes the emergence of certain other correlated states such as superconductivity. More recently, twisted bilayer graphene that is finely-tuned to a "magic angle" at which the bands flatten, has introduced a new platform for the creation of correlated phases including superconductivity[19,20].

An alternative route to flat bands that does not require fine-tuning or breaking time reversal symmetry is to create a periodic PMF by imposing a periodic strain on the 2D membrane[10]. Unlike earlier realizations of PMF which were mostly local in nature, we show here that buckling of graphene membranes produces a global change in the electronic structure resulting in a sequence of almost flat bands that percolate throughout the material.

To demonstrate the feasibility of achieving tunable flat bands, we use strain superlattices with periods ranging from 8 to 25 nm [Supplementary Information (SI), Fig. 3S], which are induced in graphene deposited onto a $NbSe_2$ substrate. We focus here on the results obtained for a superlattice with a period of 14 nm, which exhibited the characteristic behavior. Fig. 2a shows the scanning tunneling microscope (STM) topography image of a graphene crystal where the buckling results in a triangular lattice of alternating bright (crests) and dark (troughs) regions (SI Fig. 2S). The large lattice mismatch between graphene and $NbSe_2$, 0.246 nm and 0.36 nm respectively, rules out an interpretation of this structure in terms of a Moiré pattern[21-23]. We first focus on the electronic structure obtained from the *dI/dV* spectra (*I* is the current, *V* is the bias) in the crest regions (Fig. 2b). The spectra consist of a sequence of peaks that are attributed to PMF-induced pseudo Landau levels (PLL)[8-10]. The energy of the most pronounced peak (labeled *N*=0 in Fig. 2b) is aligned with the Dirac point ($E_D$). It is shifted by doping from the $NbSe_2$ substrate[24] to ~0.5eV above the Fermi energy ($E_F$) which serves as the energy origin. Labeling the remaining peaks in order of increasing (decreasing) energy as $N = \pm 1, \pm 2, ...$ we find that they follow the sequence expected of Landau levels in monolayer graphene[25,26], $E_N = E_D + sgn(N)v_F\sqrt{2e\hbar|N|B_{PMF}}$. This PLL sequence provides clear evidence of a strain induced PMF in the buckled graphene. Using the standard value of the Fermi velocity, $v_F = 1.0 \times 10^6 m/s$, we estimate the PMF in the center of the crest region as $B_{PMF} = 108 \pm 8\,T$ (Inset Fig. 2b). Outside the buckled region where the STM topography is flat, the spectrum exhibits a featureless "V" shape as expected of flat graphene on $NbSe_2$ (SI Fig. 7S). This again supports the interpretation of the observed periodic pattern in terms of buckling-induced strain rather than a moiré pattern.

One of the hallmarks of PLLs is the sublattice polarization in the $N=0$ level such that, for a given orientation of the PMF, the electronic wavefunction is confined to only one of the sublattices, say A, while in areas where the PMF orientation is reversed, the wavefunction is localized on the B sublattice. The sublattice polarization in the presence of a PMF is the counterpart of the valley polarization of the $N=0$ level in an external magnetic field and follows directly from the opposite signs of the PMF in the K and K' valleys. Thus, the sublattice polarization provides an experimental signature that reveals the presence of a PMF[27-31]. In Fig. 2d, 2e and 2f we show atomic resolution STM topography in the crest, intermediate and trough regions of the superlattice, respectively. In both crest and trough regions, the sublattice polarization is clearly revealed by the triangular structure (dashed triangles). The opposite orientations of the triangles in the two regions indicate the presence of PMFs with opposite signs. Here, the triangular structure reflects the fact that only every other atom in the honeycomb is visible, corresponding to the wavefunction being localized within one sublattice in the crest, and within the other in the trough, as illustrated by the ball and stick cartoons. In the transition regime between the crest and the trough (Fig. 2e), there is a narrow boundary where all atoms in the honeycomb are visible, indicating a zero crossing of the PMF between the two regions. By contrast, in all regions of the sample where the topography is flat, we observe the full honeycomb structure which is accompanied by a "V" shaped spectrum, in agreement with the absence of the PMF outside the buckled region (SI Fig. 7S).

The spectra in the trough regions, shown in Fig. 2c, also exhibit a sequence of peaks which are more clearly resolved in the negative second derivative of the spectrum. Plotting the peak energy versus the level index we find that the sequence is linear in $N$, instead of the square root scaling seen in the crest region, with a roughly equidistant energy spacing of ~ $94\pm5$ meV.

To better understand the electronic structure of buckled graphene, we performed tight-binding calculations in the presence of a periodically modulated PMF configuration with a triangular structure (Fig. 3d, Bottom) similar to that in the experiment (Fig. 3d, Top):

$$\boldsymbol{B_{PMF}}(x,y) = B[\cos(\boldsymbol{b_1 r}) + \cos(\boldsymbol{b_2 r}) + \cos(\boldsymbol{b_3 r})] \tag{1}$$

where $B$ is the PMF amplitude, $\boldsymbol{b_1} = \frac{2\pi}{a_b}\left(1, -\frac{1}{\sqrt{3}}\right)$, $\boldsymbol{b_2} = \frac{2\pi}{a_b}\left(0, \frac{2}{\sqrt{3}}\right)$, $\boldsymbol{b_3} = \boldsymbol{b_1} + \boldsymbol{b_2}$, and $a_b$ is the buckled superlattice period (here, $a_b$ = 14 nm). This field configuration corresponds to a periodic array of PMF crests peaked at a maximum PMF of $3B$, that are surrounded by a percolating network of troughs where the minimum PMF is $-1.5B$ (Fig. 3d, bottom). The zeros of this PMF configuration form circles that surround each crest. Because of the spatial variation of the PMF, the experimentally measured spectra correspond to an effective PMF, $B_{eff}$, which approximately averages the field over the cyclotron orbit. This averaging effect becomes more pronounced as the ratio between the magnetic length and the lattice period decreases, so that the effective $B_{eff}$ on the crest could be significantly smaller than $3B$ (Fig. 12S). Theoretically, we found that the PLL sequence obtained for $B$= 120T matches the experimentally measured sequence shown in Fig. 2b.

Figures 3a, 3b, 3c plot the evolution of the local density of states (LDOS) with the PMF, for each sublattice (A - top; B - bottom) in the crest, intermediate and trough regions respectively, marked by the symbols in Fig. 3d bottom. These figures clearly show the polarization on the A sublattice, transiting through an unpolarized state to the B sublattice polarization, consistent with the experimental results shown in Figs. 2, d-f. In Fig. 3e we show the simulated LDOS spectrum in the crest region for $B = 120T$ (top) together with a fit to the square root dependence on $N$ (bottom) which agrees with the experimentally measured spectra shown in Fig. 2b. The simulated LDOS spectrum in the trough region (Fig. 3f) is approximately linear in $N$ with a level spacing of ~ 90meV, which is also consistent with the experimental results of Fig. 2c. To elucidate this unusual linear peak sequence, we calculated the LDOS evolution (Fig. 3g) along a path connecting two crests shown by the green line in Fig. 3d bottom. The peak positions extracted from the experiment in the crest and trough regions are shown by green dashed arrows. In the center of the trough the equidistant level sequence is clearly seen. We note that these levels are not solely determined by the local value of the magnetic field and they do not exhibit spatial dispersion. Furthermore, these states are not completely localized in the trough region, but spread into the crest region. However, they do not merge with the Landau levels seen in the crest, but rather disappear as they approach it. The discrete nature of these equidistant levels indicates that they originate from strain induced confinement within the quantum well defined by the PMF, which is analogous to magnetic confinement in quantum dots in 2D semiconductor materials[32,33]. As in the case of quantum dots, here the electrons are trapped in a PMF-induced potential well, which result in a set of levels spaced by a characteristic (geometry dependent) energy scale $\Delta E \sim \frac{\hbar v_F \pi}{W}$, where *W* is the dot size. Using the energy scale of the levels in the trough region, ~94 meV, we estimate *W* ~21nm, which is approximately the size of the well indicated by the gray dashed lines in Fig. 3g. In this region, the energy of each level decreases with increasing PMF until the magnetic length becomes considerably smaller than the dot size, at which point these levels merge into one degenerate level that approaches the Dirac point, as seen in Fig. 3c.

We now discuss the emergence of flat bands in this system. The periodic potential imposed by the PMF superlattice breaks up the low energy conical band of graphene into a series of minibands whose width is controlled by the strength of the PMF amplitude, *B*. At low values of *B*, the minibands restructure the LDOS into a series of semi-discrete levels as shown in Figs. 3a-c. As *B* increases, these levels evolve into increasingly narrow bands that become flat in the limit of large *B* (Fig. 13S) where we compare the band structure and the corresponding LDOS at 140 T (SI Fig. 13S a,b) with that at 180 T (Fig. 13S c,d). In Fig. 4 we plot the first few minibands (4a), dI/dV maps (4b) and corresponding LDOS maps (4c) in the buckled graphene for *B* = 120T. All the minibands show flat-band segments along the K'-M' line in the BZ. We note that the higher the miniband order, the flatter it becomes, with the 3'rd miniband taking up almost the entire BZ. In Figs. 4b, 4c we show the DOS maps for the flat-band energy slices in three minibands, $E_0 = -0.03$eV, $E_2 = -0.17$eV, $E_3 = -0.28$ eV marked in Fig. 4a. The wave-function in the $E_0$ state is mostly localized within the crest region and, in spite of the very large DOS, it is unlikely to create conditions for a macroscopic correlated state. In contrast, in the higher order minibands, the LDOS is concentrated in the trough regions which percolate throughout the entire sample, indicating the emergence of global flat bands.

The periodic structures studied here are typically nested between wrinkles in the graphene membrane (SI Fig. 6S) which often form during sample fabrication. Topographical analysis of the wrinkles (SI Fig. 4S) suggests that the buckling was triggered by the compressive strain produced by the collapse of these wrinkles during thermal cycling. With this understanding in mind, we repeated the experiment by depositing graphene not only on NbSe$_2$ but also on hexagonal boron nitride substrates. Many of the wrinkles introduced during fabrication collapsed upon annealing (Methods) and led to the appearance of a variety of buckling patterns ranging from 1D to 2D with nanometer-scale periods (Fig.1). Similarly to folding paper in the art of origami, the shape, period and symmetry of the buckled structures can be controlled by experimentally adjustable parameters, such as boundary geometry and strain distribution, providing a flexible way for the realization of PMF-induced flat bands with prescribed geometry as a venue for exploring the effects of strong interactions and the emergence of correlated phases.

**Methods:**

To avoid oxygen and moisture contamination the heterostructures are fabricated in a dry Ar atmosphere in a glovebox. Graphene is first exfoliated on a PMMA film and then transferred onto a thin $NbSe_2$ flake that is exfoliated on an $SiO_2/Si$ substrtae. The $NbSe_2$ flake was not superconducting at the experimental temperatures. Instead of removing the PMMA sacrificial layer by acetone, we directly peeled PMMA off. The Au electrodes are deposited by standard SEM lithography. Before loading the sample into the STM chamber, the sample is annealed at 230 °C in forming gas (10% $H_2$ and 90% Ar) overnight to remove the PMMA residue. The STM experiments are performed in a home-built STM[26] at 4.6K with a cut $Pt_{0.8}Ir_{0.2}$ tip. The tip used here is calibrated on the Au electrode, and the dI/dV curves are measured using a standard lock-in technique with a small A.C. voltage modulation (2mV at 473.1Hz, except where noted) added to the D.C. junction bias[26].

# FIGURES

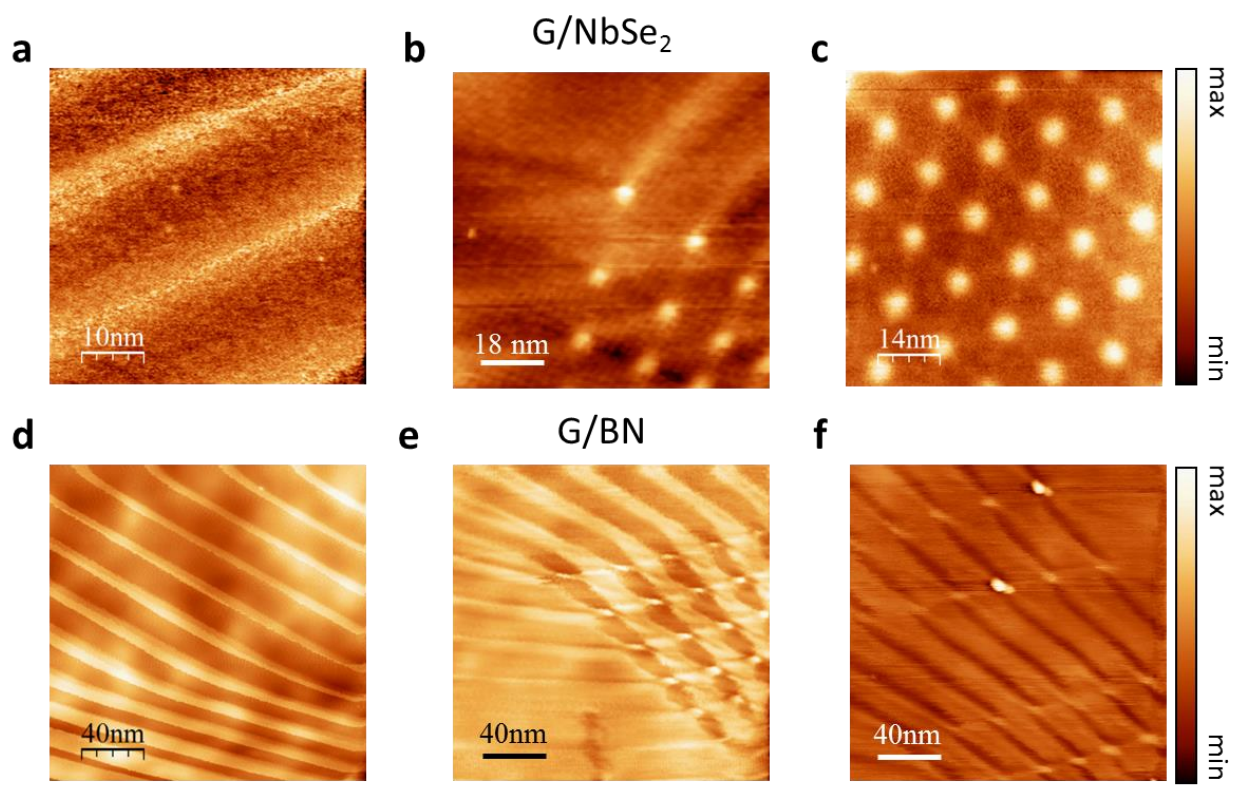

**Figure 1. Buckled structures in graphene membranes. a-c,** 1D and 2D buckling patterns observed by STM topography in G/NbSe$_2$ . **d-f**, 1D and 2D buckling modes observed by STM topography in G/hBN.

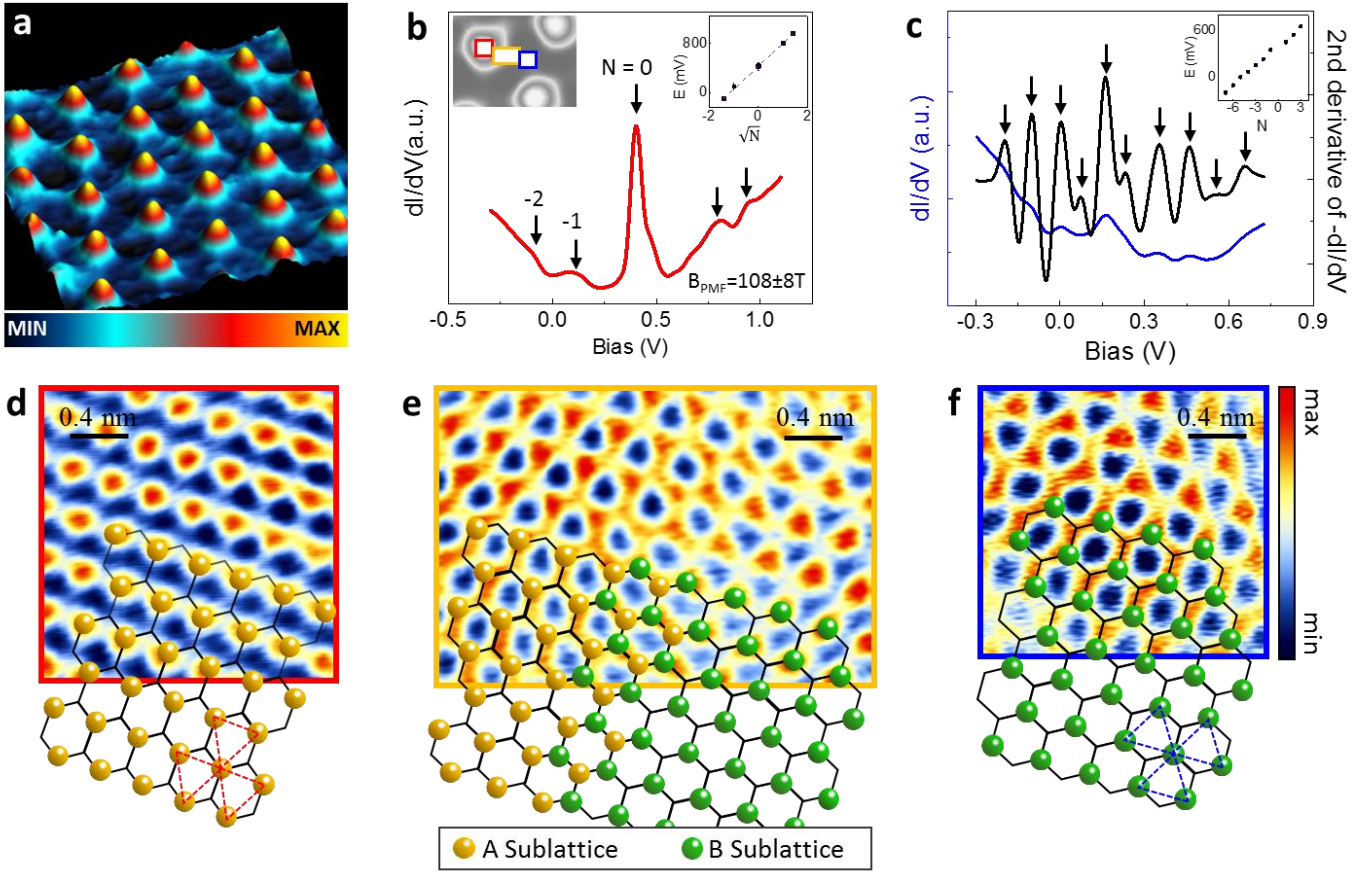

**Figure 2. Pseudo Landau level quantization and sublattice polarization in buckled graphene.**
**a,** STM topography of buckled graphene membrane supported on NbSe$_2$ reveals a 2D triangular lattice of crests with period 14 nm ($V = 600$mV, $I = 20$pA). **b,** $dI/dV$ curve on the crest area labeled by the red square in the left inset. Right inset: PLL energy plotted against the square root of the LL index, $N = 0, \pm 1, \pm 2, ....$ **c,** Same as **b** but on a trough site (blue square in the left inset of **b**). The negative second derivative of the $dI/dV$ signal is superposed (black curve) to better reveal the peaks sequence. Inset shows the approximately linear dependence of the peak energy on the level index, $N$. **d-f,** Atomic resolution STM topography in crest (red square in left inset of **b**), transition (orange rectangle) and trough (blue square) regimes (color scale shown in panel **f**). A schematic honeycomb lattice with the orange and green balls representing the two sublattices is superposed to highlight the sublattice polarization in the different regimes. The dashed-line triangles indicate the opposite orientations of the lattice polarization in each region.

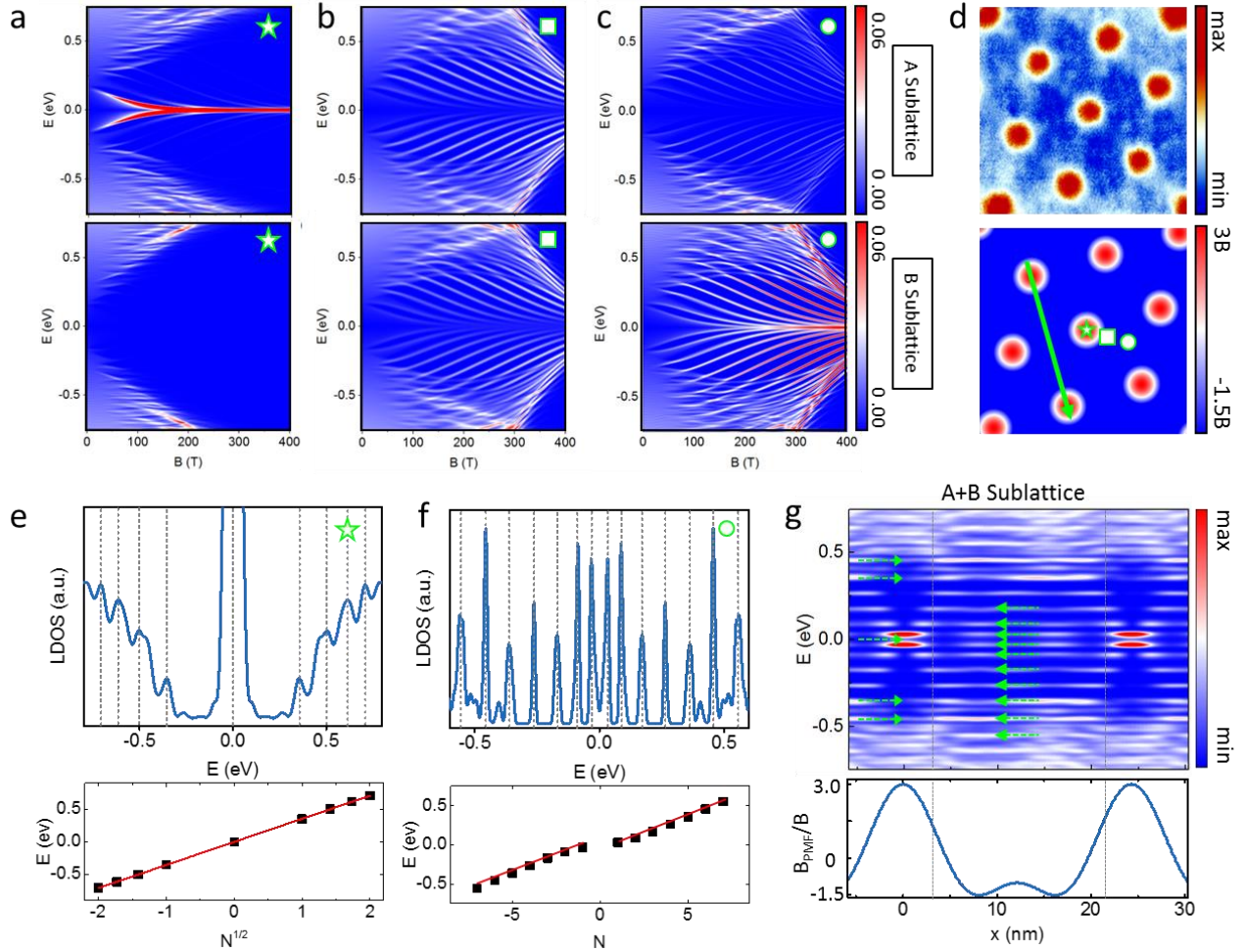

**Figure 3. Simulated LDOS in buckled graphene. a-c**, Simulated LDOS evolution with the PMF amplitude for the crest, intermediate and trough regions respectively. Top (bottom) panels represent the A (B) sublattice. **d**, Top panel: STM topography ($V = 500$mV, $I = 20$pA) showing the alternating crest and trough regions. Bottom panel: PMF configuration used in the simulation. The symbols indicated the positions of the calculated LDOS in **a-c**. **e-f**, Calculated LDOS as a function of energy in the crest and trough regions respectively for $B = 120$T. Bottom panels show the level-index dependence of the energy levels **g**, Top panel, Contour plot of the LDOS spectra (sublattice averaged) between two crest areas, along the green line in panel **d**. Green dashed arrows indicate the positions of the peaks in the measured spectra shown in Figs. 2b-c.

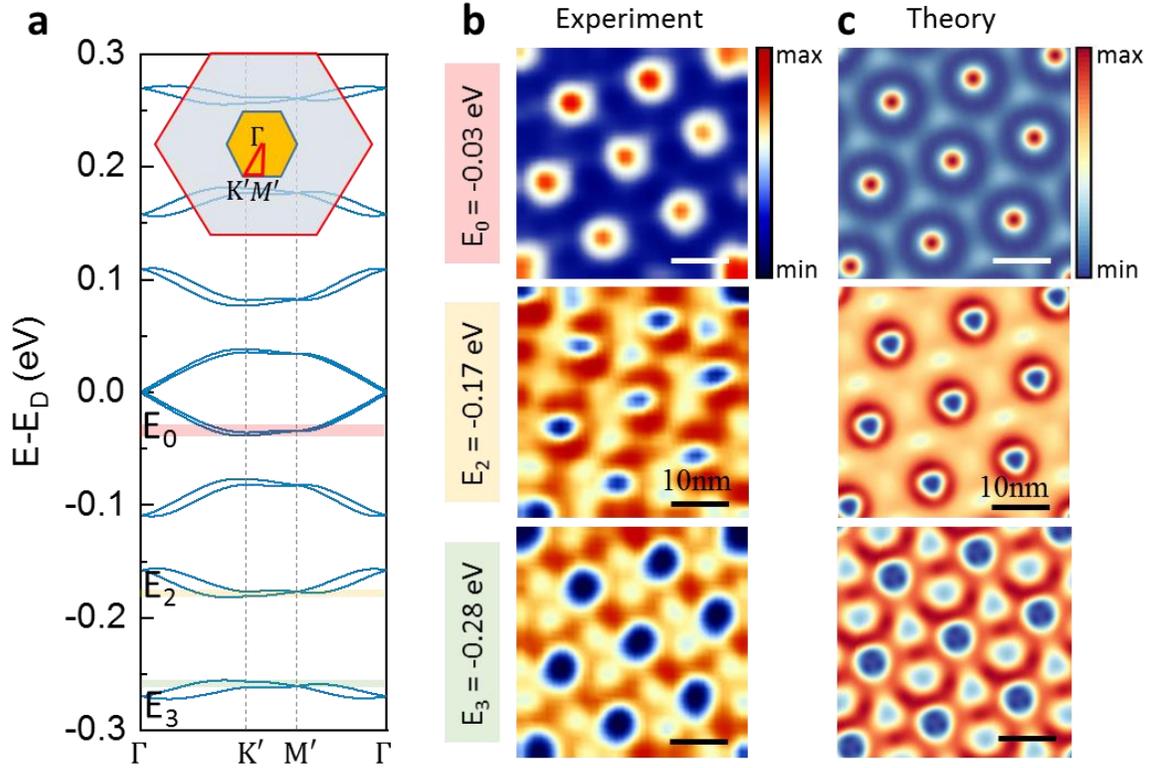

**Figure 4. Flat bands and LDOS maps. a,** Calculated band structured for buckled graphene superlattice with period 14nm and PMF amplitude $B$=120T. Inset: Superlattice mini-Brillouin zone, nested within the original Brillouin zone of flat graphene is shown together with the trajectory of along which the band structure is calculated. **b,c,** Measured $dI/dV$ maps and calculated LDOS maps at $B$=120T for the three energies that correspond to the flat band regions in panel **a** as indicated by the color coding.

# Supplementary material for

# Flat Bands in Buckled Graphene Superlattices

Yuhang Jiang[1], Miša Anđelković[2], Slaviša P. Milovanović[2], Lucian Covaci[2], Xinyuan Lai[1], Yang Cao[3, 4],

Kenji Watanabe[5], Takashi Taniguchi[5], Francois M. Peeters [2], Andre. K. Geim[3, 4], and Eva Y. Andrei[1]

[1] Department of Physics and Astronomy, Rutgers University, 136 Frelinghuysen Road, Piscataway, NJ 08855 USA

[2] Departement Fysica, Universiteit Antwerpen, Groenenborgerlaan 171, B-2020 Antwerpen, Belgium

[3] School of Physics and Astronomy, University of Manchester, Oxford Road, Manchester M13 9PL, United Kingdom

[4] Manchester Centre for Mesoscience and Nanotechnology, University of Manchester, Oxford Road, Manchester M13 9PL, United Kingdom

[5] Advanced Materials Laboratory, National Institute for Materials Science, 1-1 Namiki, Tsukuba 305-0044, Japan

This file contains:

1. **Large area STM topography of buckled graphene area and its boundaries**

2. **Compressive strain due to fold collapse**

3. **Buckling transition and pattern formation**

4. **STM topography of G/NbSe$_2$ far from the ridges**

5. **Examples of buckling patterns**

6. **Buckling patterns imaged with different bias voltages**

7. **dI/dV map on the crest area with high spatial resolution**

8. **Lattice constant dependence of pseudo magnetic field (PMF)**

9. **Effective PMF ($B_{eff}$) in the crest area**

10. **Flat bands**

## 1. Large area STM topography of buckled graphene area and its boundaries

The periodic pattern formation following a buckling transition is largely determined by the boundary conditions and stress distribution. In order to understand the buckling pattern in the G/NbSe$_2$ sample we carried out large area topography measurements that include the boundaries of the pattern (Fig. 1S, left). The triangular pattern is delimited by two intersecting ridges (~0.5µm long and 4nm tall) that form a 60° fan. Zooming into the fan area (Fig. 1S, right), we discern the buckling mode pattern. The height profile of the buckling pattern (Fig. 2S) indicates a height modulation of ~ 1.5Å. In Fig. 3S we show that the period of the pattern increases monotonically with the distance from the apex where the two ridges meet.

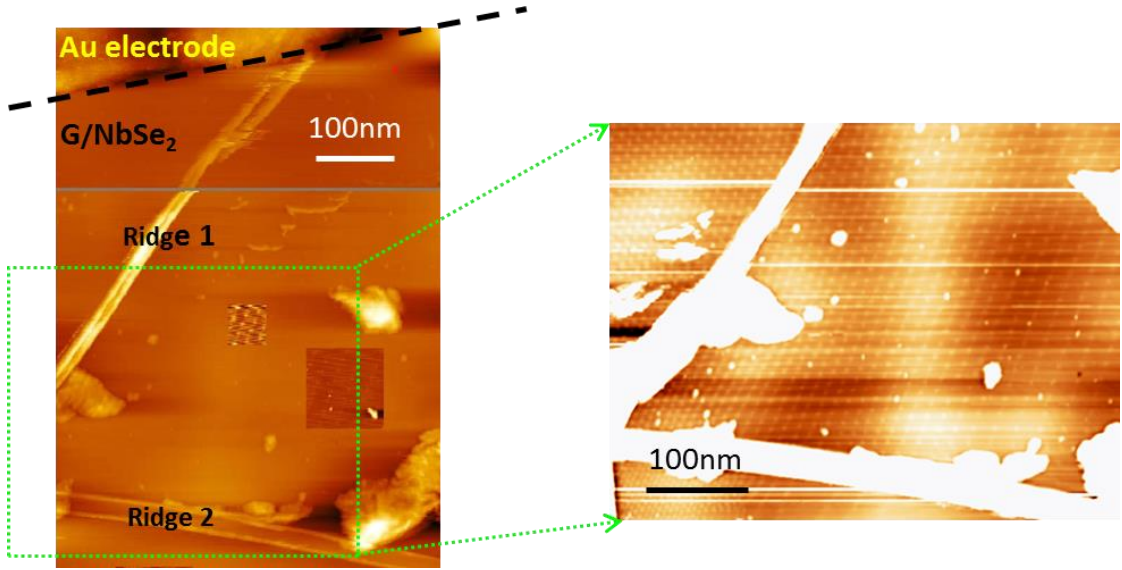

**Fig.1S Left:** Large area STM topography of G/NbSe$_2$ showing the two ridges that delimit the buckling pattern ($V_b$ = -0.3V, I = 20pA).

**Right:** Zoom in topography image showing the triangular buckling pattern ($V_b$ = 0.5V, I = 20pA).

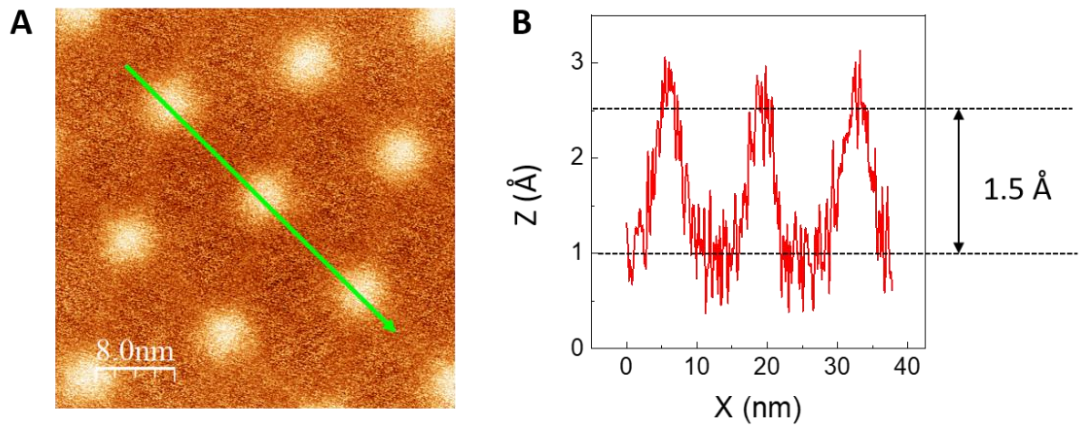

**Fig.2S** (**A**) Topography of G/NbSe$_2$ showing a high resolution buckling pattern ($V_b$ = 0.5V, I = 40pA). (**B**) Height profile along the arrow in (A) shows ~1.5Å height modulation of the buckling pattern.

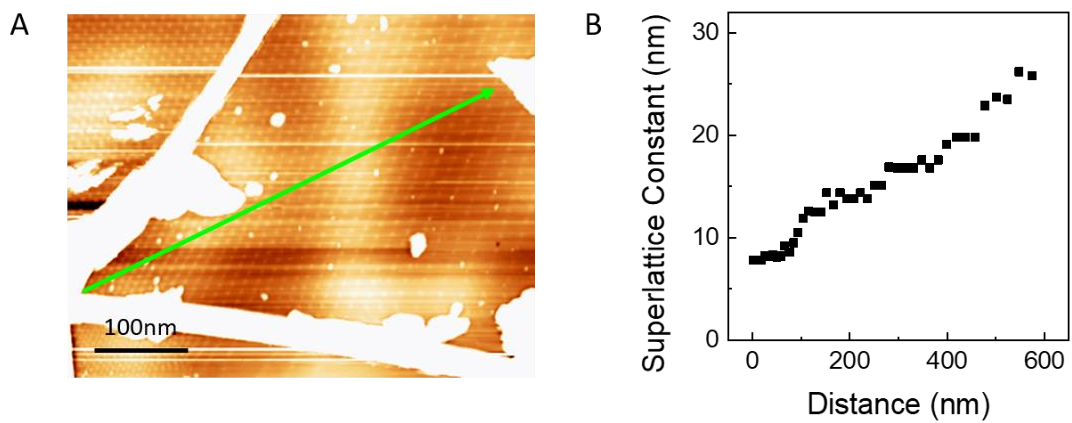

**Fig.3S** (**A**) Large area Topography of G/NbSe$_2$ ($V_b$ = 0.5V, I = 20pA). (**B**) Period of buckling pattern shows monotonically increasing period length with distance from the apex where the two ridges meet.

## 2. Compressive strain due to fold collapse

When graphene is deposited on a substrate, this process produces folds, wrinkles and bubbles due to trapped gas or solvents. Many of these defects disappear upon annealing (see methods), but not all, presumably because their geometry is such that it does not allow the trapped species to escape or because the defect is pinned to the substrate. We observe that the folds that survive the annealing step (Fig. 4S (A), (B)) do not show the usual concave profile seen prior to annealing (Fig 4S (C), top) but rather show evidence of collapse as seen by the shape in the convex area which is flanked by two tall ridges on the boundaries of the original fold (Fig 4S (C), bottom).

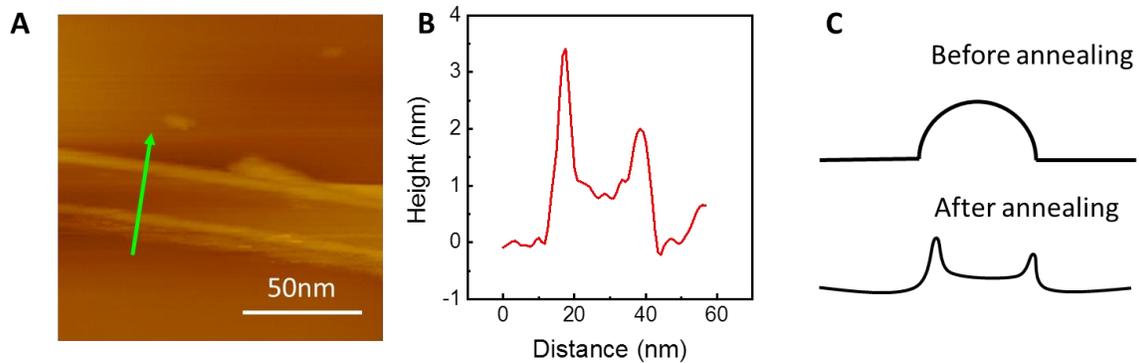

**Fig.4S** (**A**) Zoom-in of the lower ridge area in Fig. 3S (A). (**B**) Height profile of the collapsed fold along the arrow in (A). (**C**) Schematic drawing of the fold profile before and after the collapse.

We observe that the ridge pointing inwards towards the fan area is consistently shorter than the outside ridge, suggesting that some of the graphene membrane comprising the original fold was pushed inwards. This increases the area of the membrane trapped between the two ridges resulting in a compressive strain which ultimately can trigger the buckling transition. To test this scenario we conjecture that the concave region of the ridge was originally part of the convex top. This suggests that one can reconstruct the original shape by a mirror transformation of the concave part relative to the green dashed line in Fig. 5S (A) (the line intersects the tallest point of the convex part and is parallel to the substrate). This produces the reconstructed dome shown by the blue symbols in Fig. 5S (A). Using this procedure immediately reveals a missing part of the original dome of length ΔL as indicated in the figure. The strain produced by this increased length is estimated from the ratio between $\Delta L$ and the bisector of the 60° triangle: $\varepsilon = \Delta L / L \sin 30 = 2\Delta L / L$ where L is the distance from the apex formed by the intersection of the two ridges, (Fig. 5S (B)).

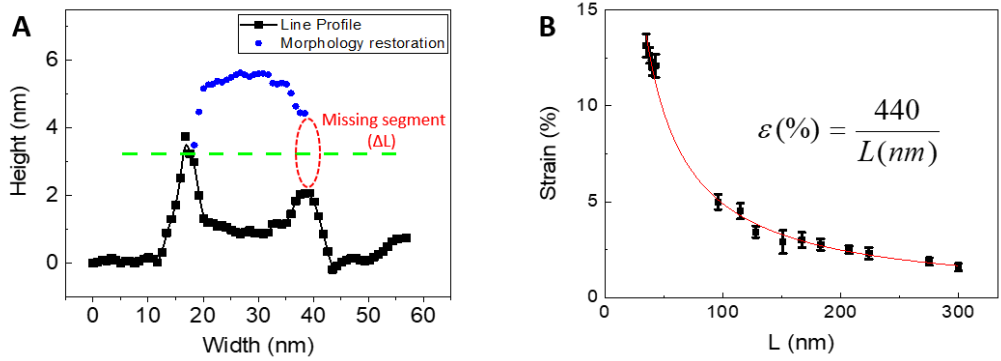

**Fig. 5S (A)** Line profile across the tallest point of the convex part. **(B)** Strain produced by the length increase calculated in (A).

### 3. Buckling transition and pattern formation

Theoretical models and simulations of wrinkling based on minimizing the energy of a stretched or compressed membrane by allowing out of plane distortions[1,2], have shown that there are simple scaling laws that relating the period of the buckled membrane $\lambda$ to the strain $\varepsilon$:

$$\frac{\lambda^4}{(tL)^2} = \frac{4\pi^2 \nu}{3(1-\nu^2)\varepsilon}.$$

Using t=0.3nm for the graphene thickness and $\nu \sim 0.15$ for its Poisson ratio and the expression for the strain as a function of distance L, we find: $\lambda = \left(\frac{4\pi^2 \nu t^2}{3(1-\nu^2)440}\right)^{1/4} L^{3/4} = 0.14 L^{3/4}$

Fitting the data for the distance dependence of the strain, shown in Fig. 3S (B) to the expression $\lambda = \lambda_0 + aL^{3/4}$ we obtain $a = 0.154 \pm 0.005$ in agreement with the estimate above. The value of the offset $\lambda_0 = (6 \pm 0.3)nm$, suggests that this formula breaks down at short distances.

To understand why the 1D scaling of the buckling period is consistent with our results we consider the sketch shown in Fig. 6S. First, we note that it is unlikely that both folds collapse at the same instant. Now if we suppose that the fold marked by the green line in the sketch collapses first, it will produce a set of roughly parallel wrinkles whose spacing increases with distance as the strain decreases according to the scaling formula above. When subsequently the yellow fold collapses it produces a similar set of wrinkles roughly parallel to itself. The points of intersection of the two wrinkle sets will thus produce a triangular pattern of crests (black dots) while the points in between will be troughs.

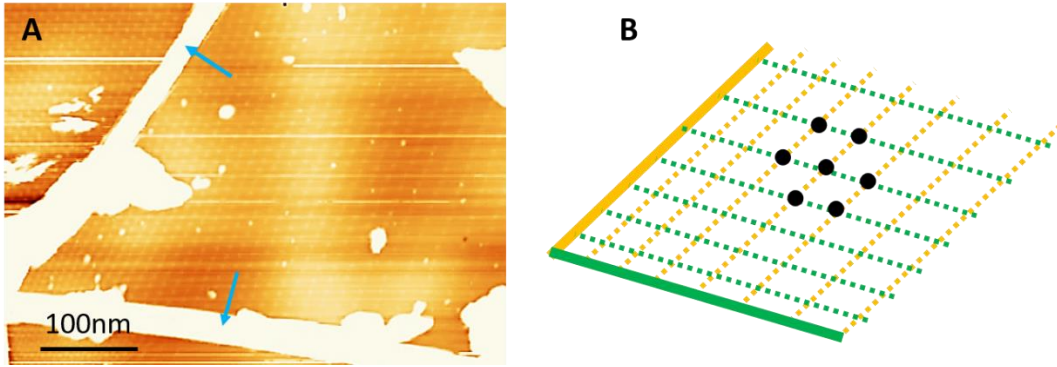

Fig.6S (**A**) Large area Topography of G/NbSe$_2$ showing the two ridges enclosing the triangular buckling pattern. (**B**) Schematics of wrinkles originating from the compressive strain at each boundary together with a sketch of the crests formed at the wrinkle intersections.

4. **STM topography of G/NbSe$_2$ for far from the ridges**

Superposing two crystal structures usually produce a periodic, moiré pattern. The maximum superperiod of such a moiré pattern between two crystals with atomic lattice constants *a* and *b* and is controlled by their lattice mismatch $\delta=(b-a)/a$, and is given by $\lambda_{max} = (1+1/\delta)a$. The lattice constants for G and NbSe$_2$ are $a = 0.246$ nm and $b = 0.36$nm, respectively, leading to $\lambda_{max} = 0.77$nm. But the observed periods of the superlattice studied here, 8nm to 24nm, immediately eliminate the possibility of interpreting the pattern in terms of a more lattice. To further confirm that the observed pattern is not due to a moiré structure we show in Fig. 7S the topography of a region far from the two ridges. The featureless STM topography (Fig.7S (A)) together with the "V" shaped dI/dV spectrum confirm that the top layer graphene is well decoupled with the bottom NbSe$_2$. Finally zoom-in image (Fig. 7S (B)) shows atomic resolution of graphene (honeycomb structure) in this region.

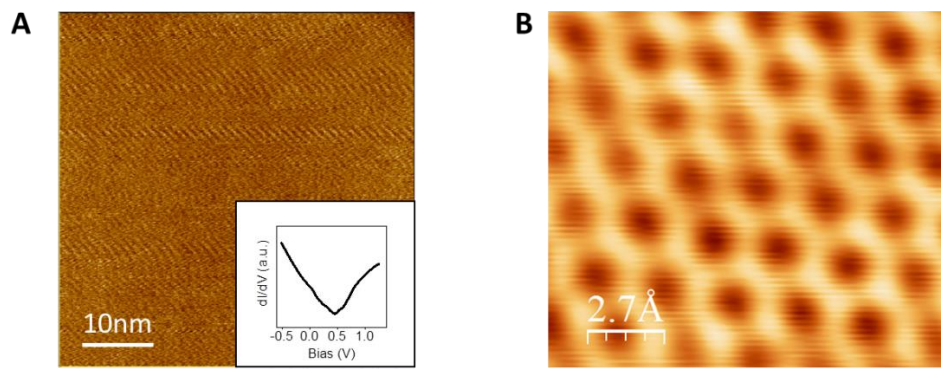

**Fig.7S** (A) STM topography of a flat G/NbSe$_2$ surface far from the ridges. $V_b$ = 0.5V, I = 30pA.

(B) Atomic resolution of G/NbSe$_2$ in (A). $V_b$ = -300mV, I = 30pA.

## 5. Examples of buckling patterns

Similar to the diversity of paper origami, the buckled graphene membranes can take on a rich variety of patterns ranging from 1D to 2D as shown in Fig.8S. The different buckling patterns are observed for both G/NbSe$_2$ and G/hBN. Interestingly, similar patterns ranging from 1D, 2D were observed in other types of membranes such as thin polymer films.

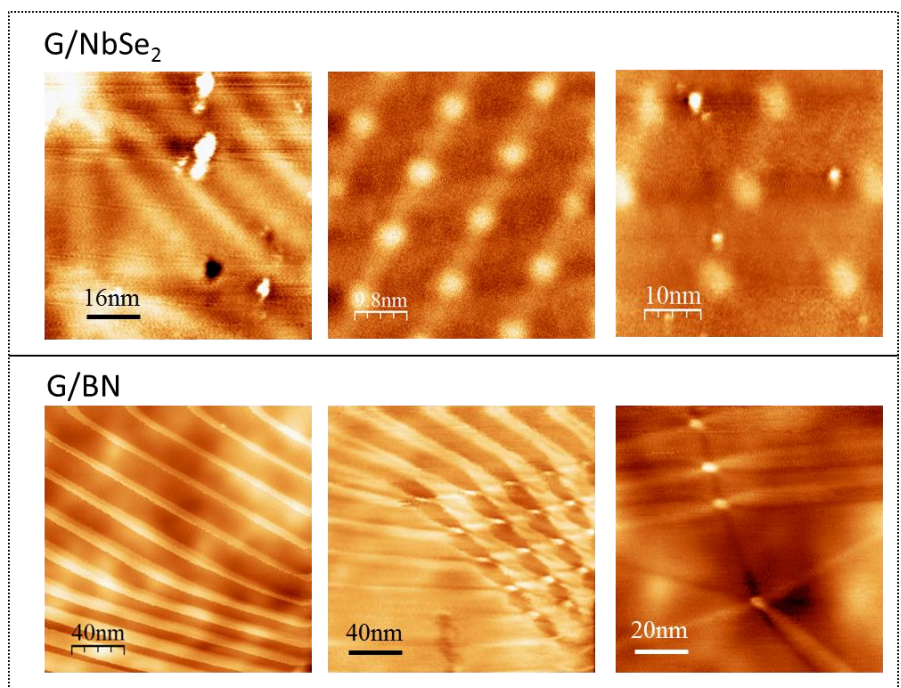

**Fig. 8S** Examples of buckling patterns of G/NbSe$_2$ (**Top**) and G/BN (**Bottom**) obtained by the mechanism discussed in the main text.

## 6. Buckling patterns imaged with different bias voltages

Fig.9S shows the STM topography of the buckled graphene taken with different bias voltages. The blue lines are guides to the eye connecting the crest areas. The crest areas remain bright for different biases consistent with their higher topography.

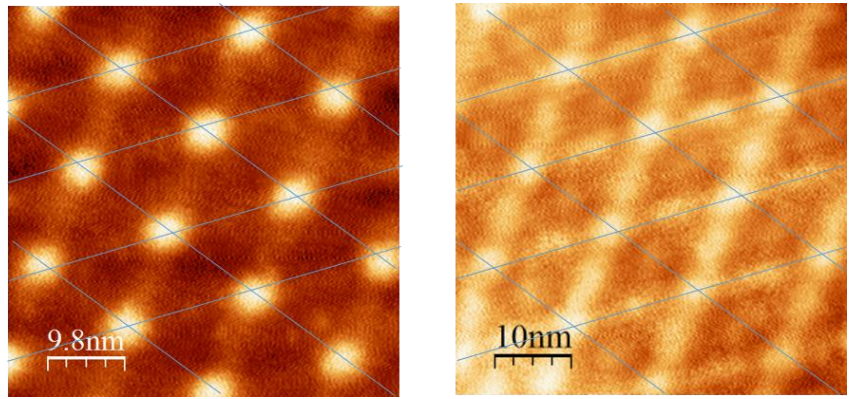

**Fig. 9S** STM topography of a region in the buckled graphene membrane measured with different bias voltages +500mV (**Left**) and +50mV (**Right**).

## 7. dI/dV map on the crest area with high spatial resolution

In Fig.10S we show the dI/dV map at an energy corresponding to the N = 0 pseudo Landau level in the crest area. The uniform LDOS represented by this map differs from the petal structure expected for a Gaussian bump and reflects the unique geometry of the Pseudomagnetic field of the induced by the buckled structure.

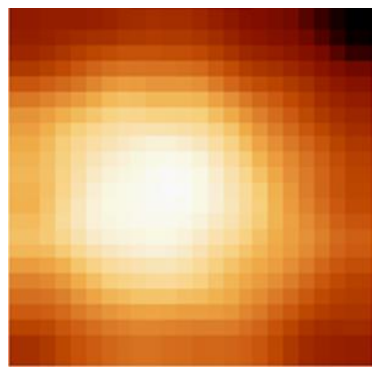

**Fig. 10S** dI/dV map over an area 6x6 nm$^2$ in the crest region at the energy of Dirac point.

## 8. Lattice constant dependence of pseudo magnetic field (PMF)

Fig.11S shows the spectra in the crest region as a function of PMF amplitude for several values of the super-period, $a_b$, as marked. The dashed lines represent the field dependence of Landau levels (LLs). We note that as the lattice spacing increases the spectra approach the uniform LL sequence at lower fields. This is consistent with the fact that LLs, which correspond to cyclotron motion, can only form if the magnetic field is fairly constant over length scales that are several times the magnetic length, $l_B = \sqrt{\frac{\hbar}{eB}} \approx \frac{25.7 nm}{\sqrt{B}}$.

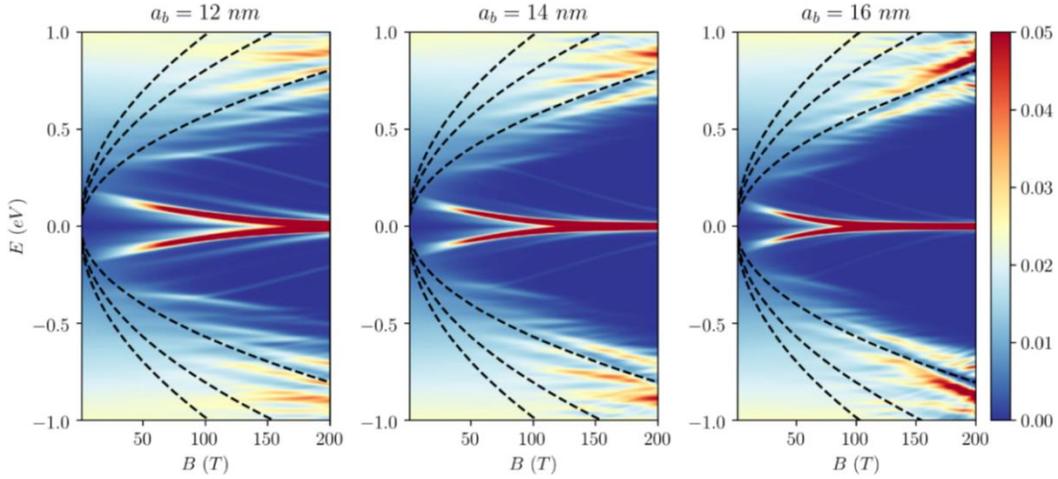

**Fig. 11S** Calculated LDOS evolution with the PMF amplitude for different supperlattice periods. Dashed lines represent the pseudo-Landau level energy dependence on field for a uniform field.

## 9. Effective PMF ($B_{eff}$) in the crest area.

In Fig.12S we show LDOS cuts from Fig.3A in the main text (A sublattice) for several values of B, as marked. For each spectrum we calculated the value of the effective magnetic field, $B_{eff}$ from the energy of the first peak $E_1 = v_F\sqrt{2e\hbar B_{eff}}$. In order to check if the peaks can be interpreted as LLs we mark by arrows the calculated peak sequence for the corresponding $B_{eff}$, $E_N = \pm v_F\sqrt{2e\hbar B_{eff} N}$. We note that the arrows coincide with the spectral peaks for sufficiently large PMF amplitudes suggesting that the Landau level language is appropriate for these spectra, as long as the $B_{eff}$ is used. However the extracted effective field is much smaller than the maximum PMF value, i.e. $B_{PMF}^{max} = 3B$, reflecting the fact that the cyclotron orbit averages the field over an area of the size $\sqrt{2N+1}l_B$. For example for a 14nm lattice period and $B = 120$T the peak sequence gives $B_{eff} = 112$T, which consistent well with the experimentally result in Fig. 2b (108±8T).

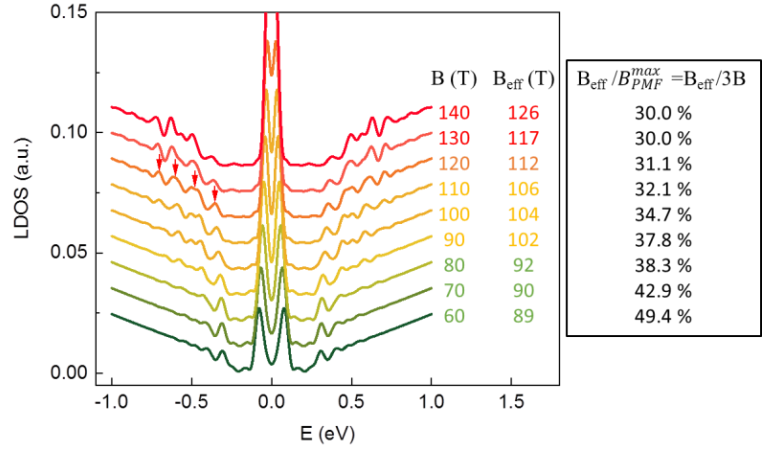

**Fig. 12S** LDOS cuts from Fig.3A (top panel) in the main text (A sublattice) for several values of B and corresponding $B_{eff}$.

## 10. Flat bands

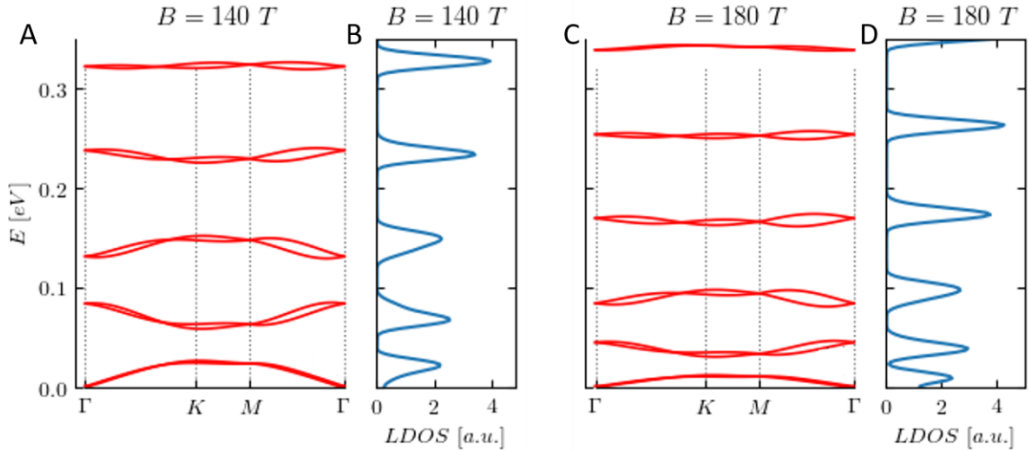

**Fig. 13S** Low energy band structure and LDOS in trough region for two values of the PMF, 140 T (A-B) and 180 T (C-D), calculated for a superlattice period of 14nm.

In Fig. 13S we plot the band structure and LDOS in the trough region for B=140T and B=180T. The figure shows the flattening of the bands with increasing field amplitude.